\documentclass[aps, prb, reprint, preprintnumbers, superscriptaddress, showpacs, amssymb]{revtex4-1}

\usepackage{graphicx}
\usepackage{dcolumn}

\begin{document}

\preprint{Ver. 8.1}¡¡

\title{Anomalous Fermi surface in FeSe seen by Shubnikov-de Haas oscillation measurements}


\author{Taichi Terashima}
\author{Naoki Kikugawa}
\affiliation{National Institute for Materials Science, Tsukuba, Ibaraki 305-0003, Japan}
\author{Andhika Kiswandhi}
\author{Eun-Sang Choi}
\author{James S. Brooks}
\affiliation{National High Magnetic Field Laboratory, Florida State University, Tallahassee, FL 32310, USA}
\author{Shigeru Kasahara}
\author{Tatsuya Watashige}
\affiliation{Department of Physics, Kyoto University, Kyoto 606-8502, Japan}
\author{Hiroaki Ikeda}
\affiliation{Department of Physics, Kyoto University, Kyoto 606-8502, Japan}
\affiliation{Department of Physical Sciences, Ritsumeikan University, Kusatsu, Shiga 525-8577, Japan}
\author{Takasada Shibauchi}
\affiliation{Department of Physics, Kyoto University, Kyoto 606-8502, Japan}
\affiliation{Department of Advanced Materials Science, University of Tokyo, Chiba 277-8561, Japan}
\author{Yuji Matsuda}
\affiliation{Department of Physics, Kyoto University, Kyoto 606-8502, Japan}
\author{Thomas Wolf}
\author{Anna E. B\"ohmer}
\author{Fr\'ed\'eric Hardy}
\author{Christoph Meingast}
\author{Hilbert v. L\"ohneysen}
\affiliation{Institute of Solid State Physics (IFP), Karlsruhe Institute of Technology, D-76021 Karlsruhe, Germany}
\author{Michi-To Suzuki}
\author{Ryotaro Arita}
\affiliation{RIKEN Centre for Emergent Matter Science, Wako 351-0198, Japan}
\author{Shinya Uji}
\affiliation{National Institute for Materials Science, Tsukuba, Ibaraki 305-0003, Japan}


\date{\today}
\begin{abstract}
We have observed Shubnikov-de Haas oscillations in FeSe.
The Fermi surface deviates significantly from predictions of band-structure calculations and most likely consists of one electron and one hole thin cylinder.
The carrier density is in the order of 0.01 carriers/ Fe, an order-of-magnitude smaller than predicted.
Effective Fermi energies as small as 3.6 meV are estimated.
These findings call for elaborate theoretical investigations incorporating both electronic correlations and orbital ordering.
\end{abstract}


\maketitle



\newcommand{\ud}{\mathrm{d}}
\def\degree{\kern-.2em\r{}\kern-.3em}

\section{Introduction}

FeSe is an intriguing material among iron-based superconductors:
The FeSe planes are isoelectronic with the (FeAs)$^{-1}$ planes of the archetypal parent compounds of iron-based superconductors such as LaFeAsO (Ref.~\onlinecite{Kamihara08JACS}) or BaFe$_2$As$_2$.  \cite{Rotter08PRL}
However, FeSe shows only a structural phase transition at $T_s\sim100$ K without an accompanying magnetic phase transition and becomes superconducting below $T_c\sim8$ K.\cite{Hsu08PNAS}  
For comparison, BaFe$_2$As$_2$  has structural and antiferromagnetic phase transitions at 140 K but does not exhibit superconductivity.\cite{Rotter08PRL}
As both transitions are suppressed by partial substitution of Ba, Fe, or As atoms, superconductivity emerges.\cite{Rotter08PRL, Sasmal08PRL}
Although the nature of the transition at $T_s$ in FeSe is not yet clear, angle-resolved photoemission spectroscopy (ARPES) measurements on FeSe have found a splitting of the $d_{xz}$ and $d_{yz}$ bands at the corner of the Brillouin zone below $\sim$110 K,\cite{Tan13NatMat, Nakayama14condmat, Shimojima14condmat} similar to one found in BaFe$_2$As$_2$,\cite{Yi11PNAS} suggesting orbital order.\cite{[{Another ARPES study proposes a different interpretation for seemingly split bands at the zone corner: }] Maletz14PRB}
Secondly, the onset temperature of superconductivity can be enhanced up to $\sim$37 K by application of pressure.\cite{Mizuguchi08APL, Medvedev09Nmat}
Moreover, it has recently been claimed that $T_c$ in single-layer FeSe films may exceed 50 K.\cite{Wang12CPL}
Finally, very recent magnetotransport, penetration depth, and spectroscopic-imaging scanning tunneling microscopy (STM) measurements on vapor-grown high-quality FeSe single crystals suggest that the Fermi energy $E_F$ is extremely small and comparable to the superconducting energy gap $\Delta$,\cite{Kasahara14PNAS} as observed previously in Te-substituted alloys Fe(Se, Te) by ARPES measurements.\cite{Lubashevsky12NPhys, Okazaki14SciRep}
FeSe may therefore be in the Bardeen-Cooper-Schrieffer (BCS)--Bose-Einstein-condensation (BEC) crossover regime.

Detailed research into the \textit{bulk} electronic structure of FeSe is necessary to advance our understanding of these intriguing properties of FeSe, but such research was impeded by difficulties in single-crystal growth.
Recently, B\"ohmer \textit{et al.} \cite{Bohmer13PRB} have grown FeSe single crystals of unprecedented quality using a vapor transport technique.
X-ray structural refinement has indicated a composition of Fe$_{0.995(4)}$Se.\cite{Bohmer13PRB}
The composition very close to stoichiometry has further been confirmed by STM topographs as well as magnetotransport data indicating a nearly perfect carrier compensation.\cite{Kasahara14PNAS}
Using those crystals, we were able to observe Shubnikov-de Haas (SdH) oscillations in FeSe.   
Our central finding is that the observed Fermi surface (FS) is extremely small and strikingly different from band-structure calculations.

\section{Experiments}

Standard four-contact resistance ($R$) measurements were performed with a 35-T resistive magnet and $^3$He or $^3$He/$^4$He dilution refrigerator at the NHMFL. 
The electrical contacts were spot-welded.
The magnetic field ($B$) direction $\theta$ is measured from the crystallographic $c$ axis.
Four samples with $T_c$ and the resistance ratio (between room temperature and 11 K) of 8.9--9.2 K and 28--32, respectively, were investigated, and consistent results were obtained.

For a purely two-dimensional FS cylinder, there would be a single SdH frequency $F$, and $F\cos\theta$ would remain constant as $\theta$ is varied.
However, in real materials, there is some $c$-axis dispersion, which modulates the cross-section of the FS cylinder.
In simple cases, two frequencies corresponding to the minimum and maximum cross-sections will appear and will exhibit upward and downward variations of $F\cos\theta$, respectively, as $|\theta|$ is increased.

\section{Results and Discussion}

\begin{figure}
\includegraphics[width=8.4cm]{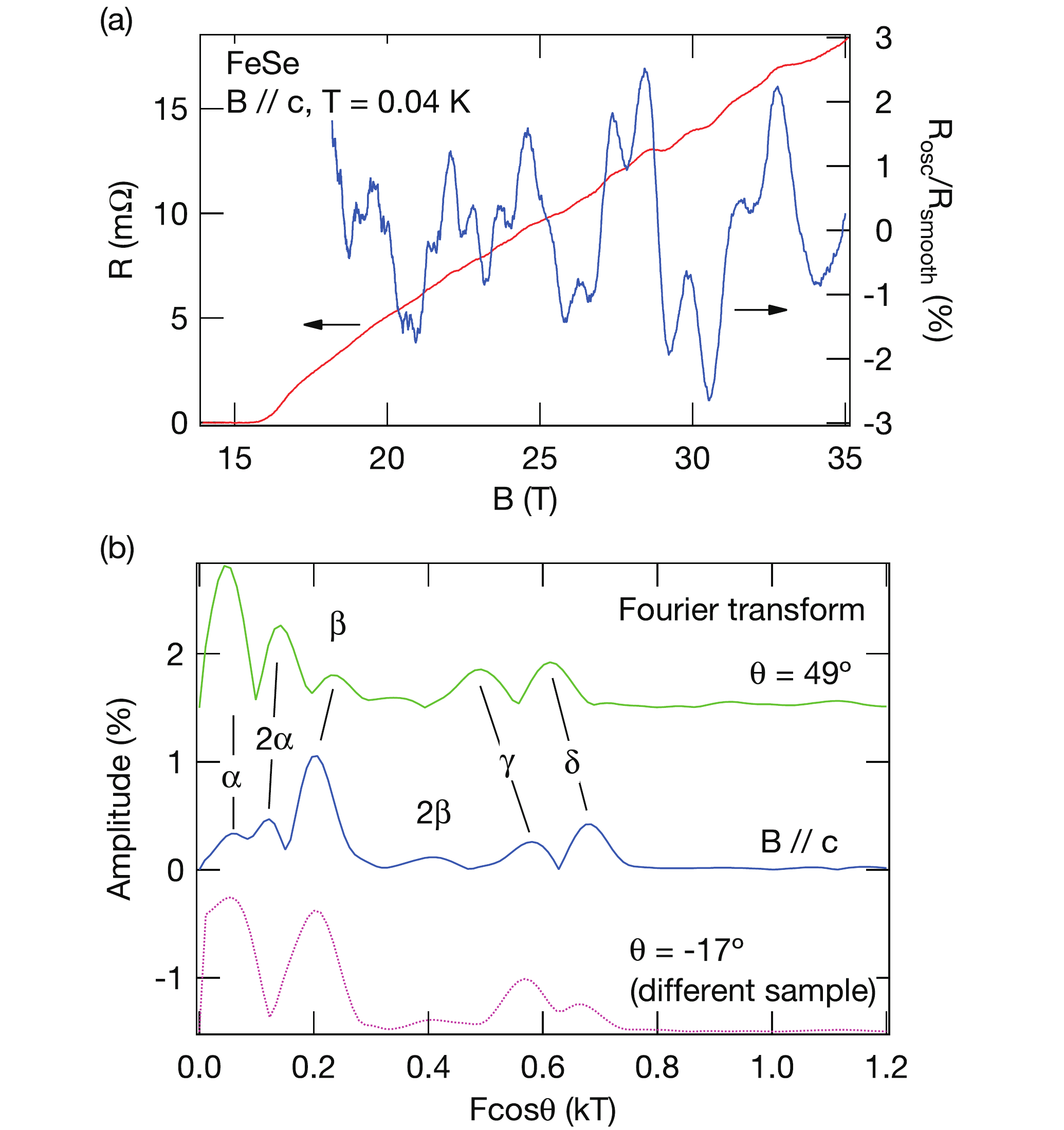}
\caption{\label{SigFT}(color online).  (a) Resistance $R$ and normalized oscillatory part $R_{osc}/R_{background}$ as a function of $B$.  A fourth-order polynomial was fitted to the former above $B$ = 18 T and was subtracted from it to obtain $R_{osc}$.  (b) Fourier transforms of SdH oscillations in inverse field vs. $F\cos\theta$. Spectra for $\theta = 0$ and 49$^{\circ}$ were taken for sample 2, for -17$^{\circ}$ for sample 3.  $T$ = 0.04 K.}   
\end{figure}

\begin{figure}
\includegraphics[width=8.4cm]{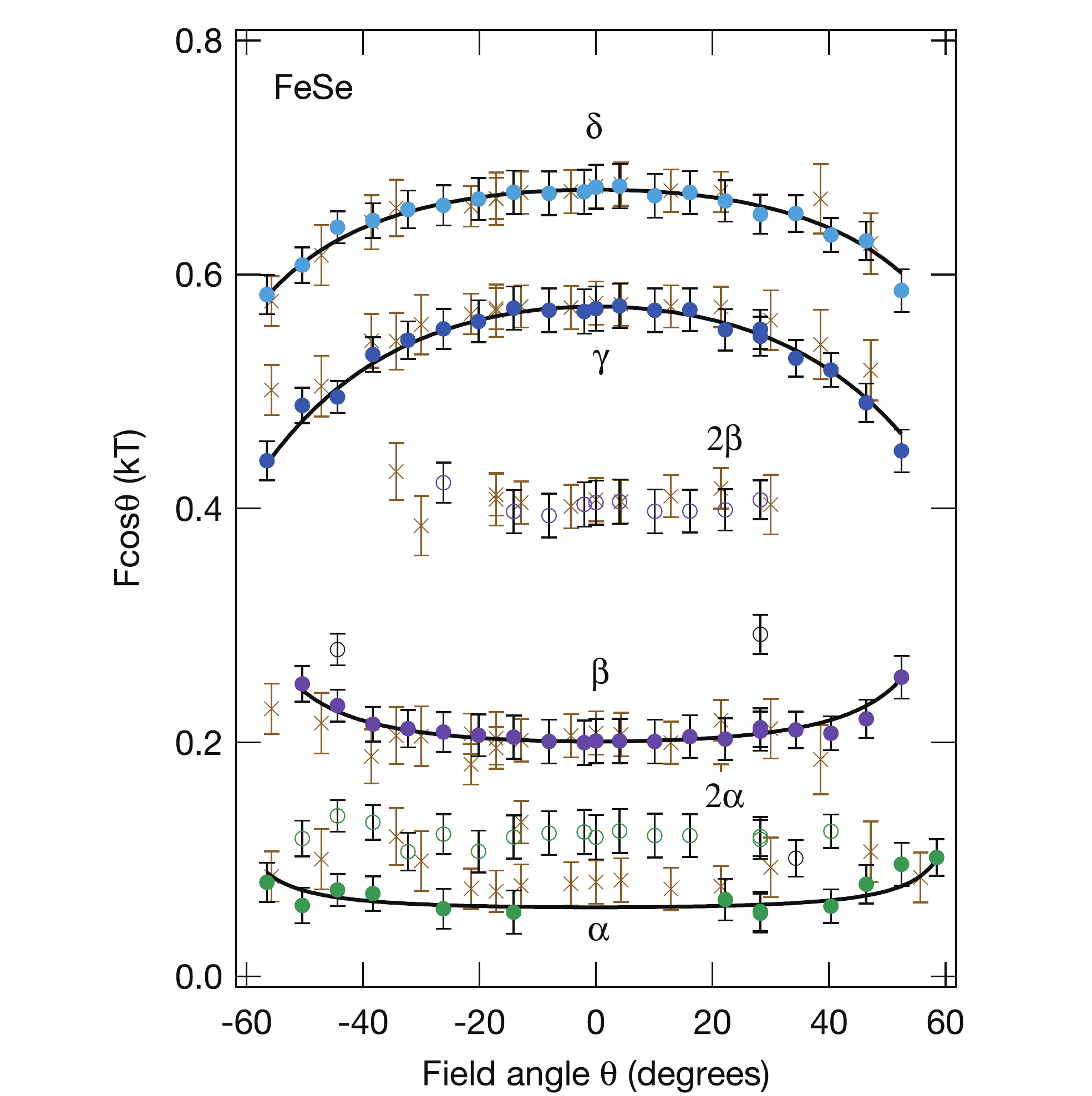}
\caption{\label{Fcos_vs_Ang}(color online).  Angle dependences of the SdH frequencies.  The vertical axis is $F\cos\theta$.  The circles and crosses are data for sample 2 ($T$ = 0.4 K) and 3 ($T$ = 0.04 K), respectively.  For the former, different frequency branches are indicated by different colors, and harmonics are indicated by hollow marks.  The solid curves are hyperboloidal- and ellipsoidal-surface fits to $\alpha$ and $\beta$, and $\gamma$ and $\delta$ in sample 2, respectively (see Appendix B).
}   
\end{figure}

\begin{table}
\caption{\label{Tab1}Experimental SdH frequencies, effective masses, orbit areas $A$, Fermi momentums and effective Fermi energies in FeSe for $B \parallel c$.  The values were averaged over the four samples except for  the $\alpha$ branch, for which the values are based on the second-harmonic data of sample 2.  $m_e$ is the free electron mass.}
\begin{ruledtabular}
\begin{tabular}{cccccc}
Branch & $F$ (kT) & $m^*/m_e$ & $A$ (\%BZ) & $k_F$ (\AA$^{-1}$) & $E_F$ (meV)\\
\hline
$\alpha$ & 0.06 & 1.9(2) & 0.20 & 0.043 & 3.6\\
$\beta$ & 0.20 & 4.3(1) & 0.69 & 0.078 & 5.4\\
$\gamma$ & 0.57 & 7.2(2) & 2.0 & 0.13 & 9.1\\
$\delta$ & 0.68 & 4.2(2) & 2.3 & 0.14 & 18\\
\end{tabular}
\end{ruledtabular}
\end{table}

\begin{figure}
\includegraphics[width=8.4cm]{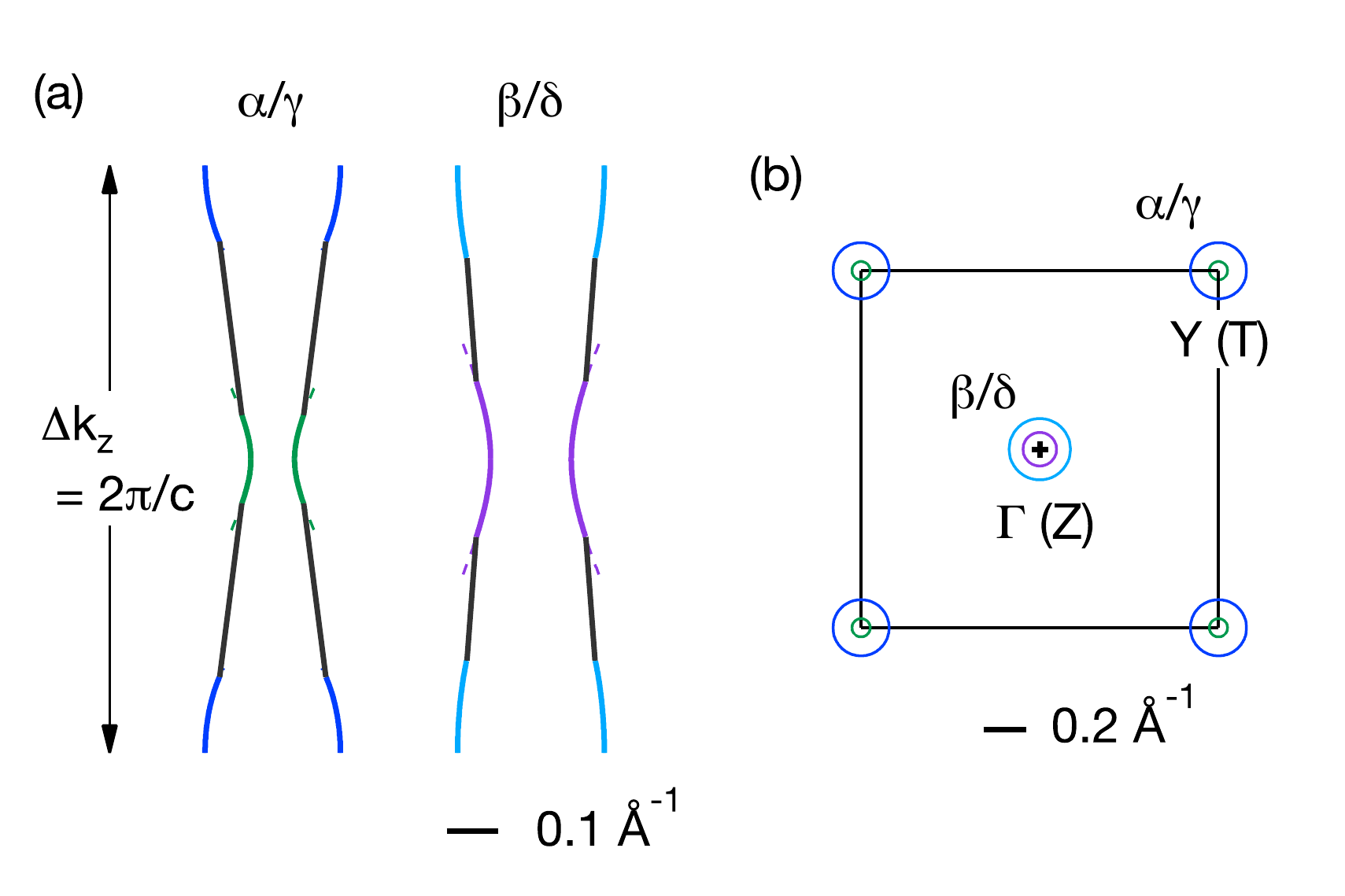}
\caption{\label{FSmodel2}(color online).  Experimental Fermi surface cross-sections containing the $k_z$ direction (a) and perpendicular to it (b).  The in-plane anisotropy is ignored, and (b) is based on the second scenario (see text).  The color coding is the same as that in Fig.~\ref{Fcos_vs_Ang} to show from which frequency branch each part of the cylinders is determined.  The black lines in (a) indicate connecting regions between hyperboloidal and ellipsoidal ones. 
}   
\end{figure}

\begin{figure}
\includegraphics[width=8.6cm]{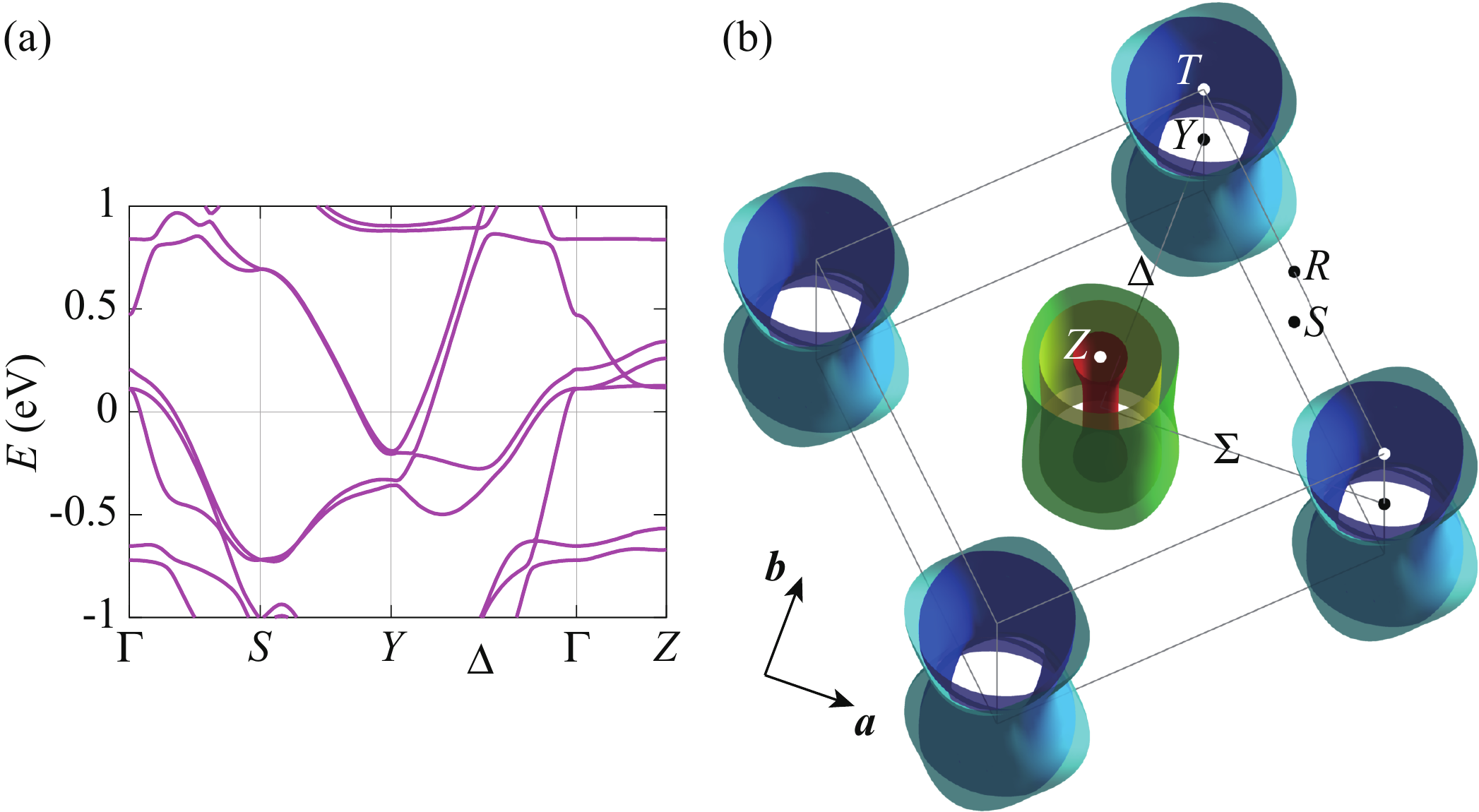}
\caption{\label{Band_calc}(color online).  Calculated band structure (a) and Fermi surface (b) of FeSe in the orthorhombic structure.  The $\Gamma$ point is the center of the Brillouin zone.  The points Y, T, S, and R correspond to M, A, X, and R of the tetragonal Brillouin zone, respectively.}   
\end{figure}

\begin{figure}
\includegraphics[width=8.4cm]{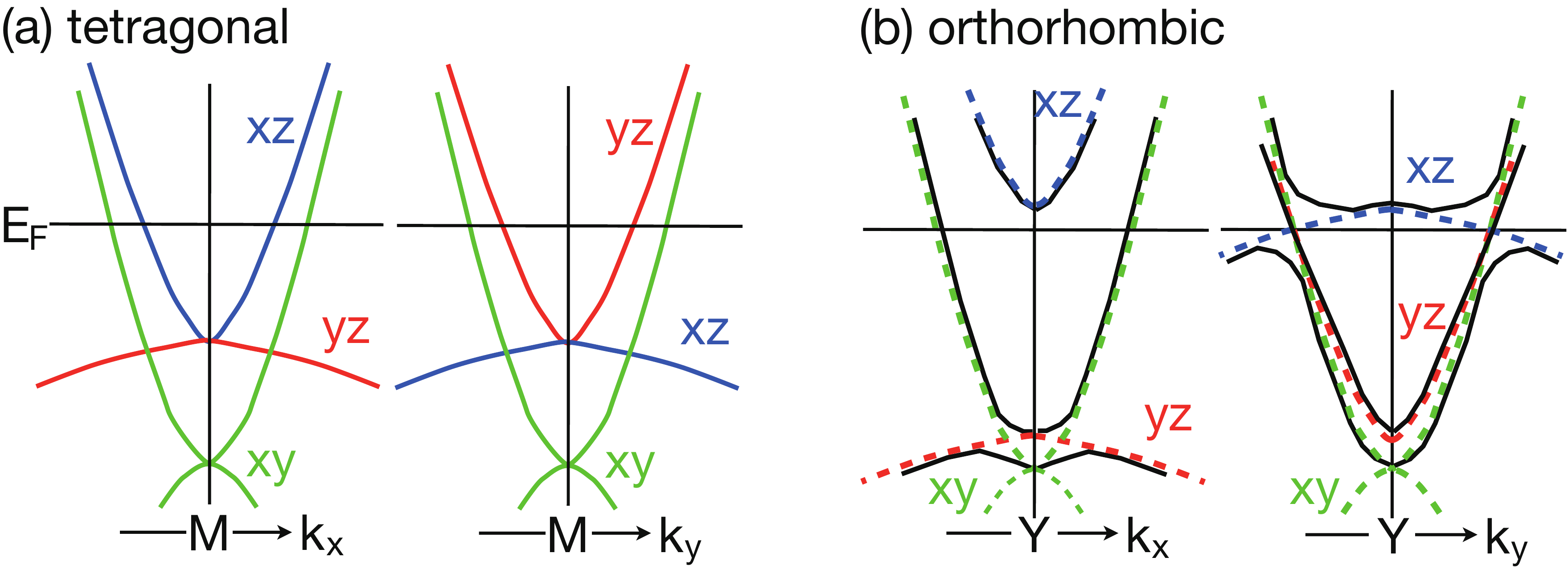}
\caption{\label{single_e_cylinder}(color online).  Schematic of band reconstruction due to the orbital order.  $k_x \parallel a$ and $k_y \parallel b$.  (a) Band structure near the M points in the tetragonal phase.  The two M points along $k_x$ and $k_y$ are equivalent except that the $d_{xz}$ and $d_{yz}$ orbital characters are inverted.  (b) Band structure near the Y points in the orthorhombic phase.  Because of the orbital order, the $d_{xz}$ band is shifted up, while $d_{yz}$ down as indicated by the broken lines.  Since the bands anticross, they are eventually reconstructed as indicated by the solid lines, resulting in a single electron cylinder.  (Note that, because of the choice $a < b$ in the orthorhombic phase, the shifts of the $d_{xz}$ and $d_{yz}$ bands are reversed in comparison to some previous works.\cite{Tan13NatMat, Nakayama14condmat, Shimojima14condmat, Yi11PNAS, Zhang12PRB})}   
\end{figure}


Figure~\ref{SigFT}(a) shows the resistance at $T$ = 0.04 K in sample 2 as a function of $B$ applied parallel to the $c$ axis.
After subtracting a smoothly varying background, we see clear SdH oscillations.
Figure~\ref{SigFT}(b) shows Fourier transforms of the oscillations vs. $F\cos \theta$ for three field directions.
The upper two spectra were obtained for sample 2, while the bottom one for sample 3. 
We find six frequency branches, $\alpha$, 2$\alpha$, $\beta$, 2$\beta$, $\gamma$, and $\delta$.
They are all small, and the corresponding orbits cover only 0.2--2.3\% of the Brillouin zone [Table I and Fig.~\ref{FSmodel2}(b)].
Figure~\ref{Fcos_vs_Ang} shows the angle dependences of the SdH frequencies for samples 2 and 3.
Note that the vertical axis is $F\cos\theta$.
The two samples show consistent angle dependences.
Although the data for sample 2 were those at $T$ = 0.4 K because the angular variation was investigated more thoroughly in the $^3$He refrigerator, no new frequency was found in additional measurements on this sample at $T$ = 0.04 K.
Within experimental accuracy, $F_{2\alpha} = 2F_{\alpha}$, and $F_{2\beta} = 2F_{\beta}$, indicating that the $2\alpha$ and $2\beta$ frequencies are the second harmonics (we have also confirmed that $m^*_{2\beta}=2m^*_{\beta}$ for $B \parallel c$ in sample 2).
We have determined effective masses $m^*$ for $B \parallel c$ from the temperature dependences of the oscillation amplitudes as tabulated in Table I.
Mean free paths $l$ can be estimated only roughly because of the limited range of inverse field. We find $l$ $\sim$30 and 80 nm for the $\beta$ and $\delta$ orbits in sample 2, respectively.

We first consider the issue of the BCS-BEC crossover.
An effective Fermi energy $E_F$ can be estimated from experimental values of $F$ and $m^*$ using the following formulae:
$E_F= \hbar^2 k_F^2/(2m^*)$, $A=\pi k_F^2$, and $F=\hbar A/(2\pi e)$, where $A$ is the orbit area in the $k$ space and we have assumed circular orbits.
The estimated Fermi energies are very small (Table I).
Hence the ratio $k_BT_c/E_F$ is large, ranging between 0.04 ($\delta$) and 0.22 ($\alpha$).
The proximity to the crossover may also be assessed by the parameter $(\xi k_F)^{-1}$ corresponding to $\sim \Delta/E_F$.\cite{Chen05PhysRep, Sensarma06PRL}
Using $\xi$ = 5.7 nm (see Appendix A for the upper critical field and coherence length) and the estimated $k_F$ values (Table I), $(\xi k_F)^{-1}$ = 0.13 ($\delta$) and 0.41 ($\alpha$).
Since the BCS theory is based on the relation that $k_BT_c \sim \Delta \ll E_F$, these estimates suggest that the superconductivity in FeSe might not fully be understood within the BCS framework.
Thus it seems worth looking for possible manifestations of the BCS-BEC crossover in FeSe, though they may substantially differ from those expected for single-band superconductors.

We now switch to the Fermi surface.
The angular dependences in Fig.~\ref{Fcos_vs_Ang} indicate that the $\alpha$ and $\beta$ frequencies are from minimal cross-sections, while $\gamma$ and $\delta$ are from maximal ones.
The former can be described by hyperboloidal surfaces while the latter by ellipsoidal surfaces as indicated by the solid curves in Fig.~\ref{Fcos_vs_Ang} (see Appendix B for details of the fits).

Based on these fits, we model the observed FS cylinders as shown in Fig.~\ref{FSmodel2}.                                
We attribute $\alpha$ and $\gamma$ to one cylinder and $\beta$ and $\delta$ to another.
This is the only reasonable combination: if $\beta$ and $\gamma$ were paired, extra minimum and maximum cross-sections would occur.  
Each cylinder has hyperboloidal, ellipsoidal, and connecting regions.
We assume that the range of $k_z$ for both hyperboloidal and ellipsoidal  regions is restricted by that covered by orbits for $\theta=55^{\circ}$.
For larger $|\theta|$, orbits enter the cone-shaped connecting region outside this $k_z$ range.
The $\alpha$/$\gamma$ cylinder contains 0.0093 carriers/Fe and the $\beta$/$\delta$ 0.015 carriers/Fe.
We can also estimate contributions of the observed FS cylinders to the Sommerfeld coefficient.
Using the average of the effective masses for the minimum and maximum orbits for each cylinder with a two-dimensional approximation, we obtain 3.2 and 3.0 mJ/molK$^2$ for the $\alpha$/$\gamma$ and $\beta$/$\delta$ cylinders, respectively.
Since the effective masses for the $\alpha$ and $\gamma$ orbits differ considerably, the former may not be very accurate, and, if the large effective mass is restricted to the swollen region of the cylinder near the $\gamma$ orbit, it may be an overestimate.

For the sake of comparison, we have performed band-structure calculations for the orthorhombic structure using the \textsc{WIEN2K} code \cite{WIEN2K} as shown in Fig.~\ref{Band_calc}.
The used lattice parameters are $a$ = 5.3078 \AA, $b$ = 5.3342 \AA, $c$ = 5.486 \AA,\cite{Margadonna08ChemComm} and $z_{\mathrm{Se}}$ = 0.266689.\cite{Bohmer13PRB}
The calculated FS consists of two electron cylinders at the Y point of the Brillouin zone and three hole cylinders at the $\Gamma$ point, similar to the iron-pnictide parent compounds.
The calculated carrier density and Sommerfeld coefficient are $n_e=n_h$ = 0.17 carriers/Fe  and $\gamma_{band}$ = 4.6 mJ/mol K$^2$.

The question now is: are the two observed cylinders electrons or holes ?
Quantum oscillation measurements on the iron-pnictide parent compounds so far indicate that electron surfaces are generally easier to observe.\cite{Coldea08PRL, Analytis09PRL, Shishido10PRL, Analytis10PRL, Putzke12PRL}
It is thus tempting to assign the observed cylinders to the two calculated electron ones.
However, considerations on the Sommerfeld coefficient are unfavorable to this scenario.
A previous specific-heat measurement on a single crystal of FeSe$_{0.963}$ in magnetic fields up to 9 T reported a Sommerfeld coefficient of 5.73$\pm$0.13 mJ/molK$^2$.\cite{Lin11PRB}
A recent measurement on vapor-grown FeSe at $B$ = 14 T has found a similar value ($\sim$5.9 mJ/molK$^2$).\cite{Hardy14unpub}
On the other hand, the sum of the above estimated coefficients for the observed cylinders is already 6.2 mJ/molK$^2$.
Further, at least one unobserved hole cylinder would have to exist in this scenario to satisfy the carrier compensation, and effective masses for the hole cylinder would most likely be no smaller than those for the electron ones (otherwise oscillations from the hole cylinder would have been detected).
Hence the total would become still larger and be difficult to reconcile with the specific-heat data.

The above considerations lead us to assume that we have observed the whole Fermi surface consisting of an electron and a hole cylinder.
We may assign the $\alpha$/$\gamma$ cylinder to electrons ($n_e$ =  0.0093 carriers/Fe) and $\beta$/$\delta$ to holes ($n_h$ = 0.015 carriers/Fe).
Then, the small carrier imbalance is consistent with the Fe-deficient composition within error.
As shown below, this second scenario means radical changes to the calculated band structure, but it can be reconciled with reported ARPES data.

We first consider the holes.
Inspection of the calculated band structure along the $\Gamma Z$ line [Fig.~\ref{Band_calc}(a)] suggests that, because of the $k_z$ dispersion of bands, it is difficult to have a single hole cylinder at $\Gamma$ by simple constant band-energy shifts.
On the other hand, ARPES measurements on FeSe indicate that only one hole band crosses the Fermi level at low temperatures to form a single hole sheet at $\Gamma Z$.\cite{Maletz14PRB, Nakayama14condmat, Shimojima14condmat}
Further, Ref. \onlinecite{Maletz14PRB} suggests that the $k_z$ dispersion of the hole band along the $\Gamma Z$ line is $\sim$10 meV.
This is consistent with our $\beta$/$\delta$ cylinder, for which the $k_z$ dispersion can be estimated from the difference in the effective Fermi energies of the $\beta$ and $\delta$ orbits to be 13 meV.
Strictly, Ref. \onlinecite{Maletz14PRB} claims that the hole band sinks below $E_F$ in parts of the $\Gamma Z$ line to form a closed pocket rather than a cylinder.
However, this discrepancy could be attributed to surface effects such as surface band-bending.\cite{Analytis10PRB}

We next turn to the electrons.
In the tetragonal structure, if the spin-orbit coupling is neglected, the two electron bands responsible for the electron cylinders are degenerate by symmetry along the MX and AR lines, which correspond to the YS and TR lines of the orthorhombic Brillouin zone.
Even if the spin-orbit coupling and tiny orthorhombic distortion ($|a-b|/(a+b) \sim 2 \times 10^{-3}$)\cite{Bohmer13PRB} are included in band-structure calculations, they remain quasi-degenerate along these lines and produce two electron cylinders as shown in Fig.~\ref{Band_calc}.
On the other hand, if we take the splitting of the $d_{xz}$ and $d_{yz}$ bands observed in ARPES measurements \cite{Tan13NatMat, Nakayama14condmat, Shimojima14condmat} and anticrossing of bands into account, it is possible to have a single electron cylinder at the zone corner as illustrated in Fig.~\ref{single_e_cylinder}.
Note however that this figure is very conceptual and that realistic models would have to include band renormalization and shifts, especially those of the $d_{xy}$ band.
Although the ARPES papers on FeSe do not state whether there is a single electron cylinder or two, there is an ARPES study on NaFeAs which shows that, while two electron cylinders exist at the zone corner in the tetragonal phase, only one exists in the orthorhombic phase due to the electronic reconstruction at $T_s$.\cite{Zhang12PRB}

We now discuss a remarkable disparity between the calculated and observed carrier densities: $n_e=n_h$ = 0.17 carriers/Fe vs. $n_e$ =  0.0093 and $n_h$ = 0.015 carriers/Fe.
In iron-based superconductors, upward and downward shifts of electron and hole bands, respectively, relative to band structure calculations are often found, resulting in smaller Fermi surfaces.\cite{Coldea08PRL, Analytis09PRL, Shishido10PRL, Analytis10PRL, Putzke12PRL, Yi09PRB}
This FS shrinking has been attributed to electronic correlation effects, especially interband scattering.\cite{Zhai09PRB, Ortenzi09PRL, Aichhorn10PRB}
For example, the Fermi surface of BaFe$_2$(As$_{1-x}$P$_{x}$)$_2$ shrinks as $x$ is decreased from 1, while the mass enhancement, a measure of the correlations, and $T_c$ increase.\cite{Shishido10PRL, Analytis10PRL}
The carrier density at $x$ = 0.63 is 0.05 carriers/Fe.\cite{Analytis10PRL}
At $x$ = 0.41, where $T_c \sim 25$ K, the Fermi surface is roughly twice smaller than calculated.\cite{Shishido10PRL}
However, the magnitude of the present shrinking is the largest ever observed.
It is interesting to note that the observed carrier density is fairly comparable to that in the antiferromagnetic state of BaFe$_2$As$_2$  ($n_e=n_h$ = 0.006 carriers/Fe),\cite{Terashima11PRL} where most of the paramagnetic FS has been destroyed by the reconstruction at the antiferromagnetic transition.
There are some theoretical works on the electronic structure of FeSe where the dynamical mean-field theory (DMFT) \cite{Aichhorn10PRB, Mandal14PRB} or GW approximation \cite{Tomczak12PRL} is used to treat the electronic correlations beyond the level of conventional band-structure calculations.
They predict slightly modified Fermi surfaces compared to conventional calculations but can not explain our extremely small Fermi surface.


In conclusion, we have observed SdH oscillations in FeSe.
Our analyses indicate that the Fermi surface in the orthorhombic state is very different from that expected from band-structure calculations, most likely consisting of one hole and one electron tiny cylinders.
To elucidate how this radical deviation occurs is an urgent task, when effects of both the electronic correlations and the orbital order have to be considered.
It will be very interesting to see how this anomalous Fermi surface evolves as $T_c$ increases with pressure.
 
\begin{acknowledgments}
This work has been supported by Japan-Germany Research Cooperative Program, KAKENHI from JSPS and Project No. 56393598 from DAAD, and the ¡ÈTopological Quantum Phenomena¡É (No. 25103713) KAKENHI on Innovative Areas from MEXT of Japan.
A portion of this work was performed at the NHMFL, supported by NSF Cooperative Agreement DMR-1157490, the State of Florida, and the US DoE.
JSB acknowledges support from NSF-DMR 1309146.
\end{acknowledgments}

\appendix
\section{Upper critical field $B_{c2}$ and coherence length $\xi$}

\begin{figure}
\includegraphics[width=8.4cm]{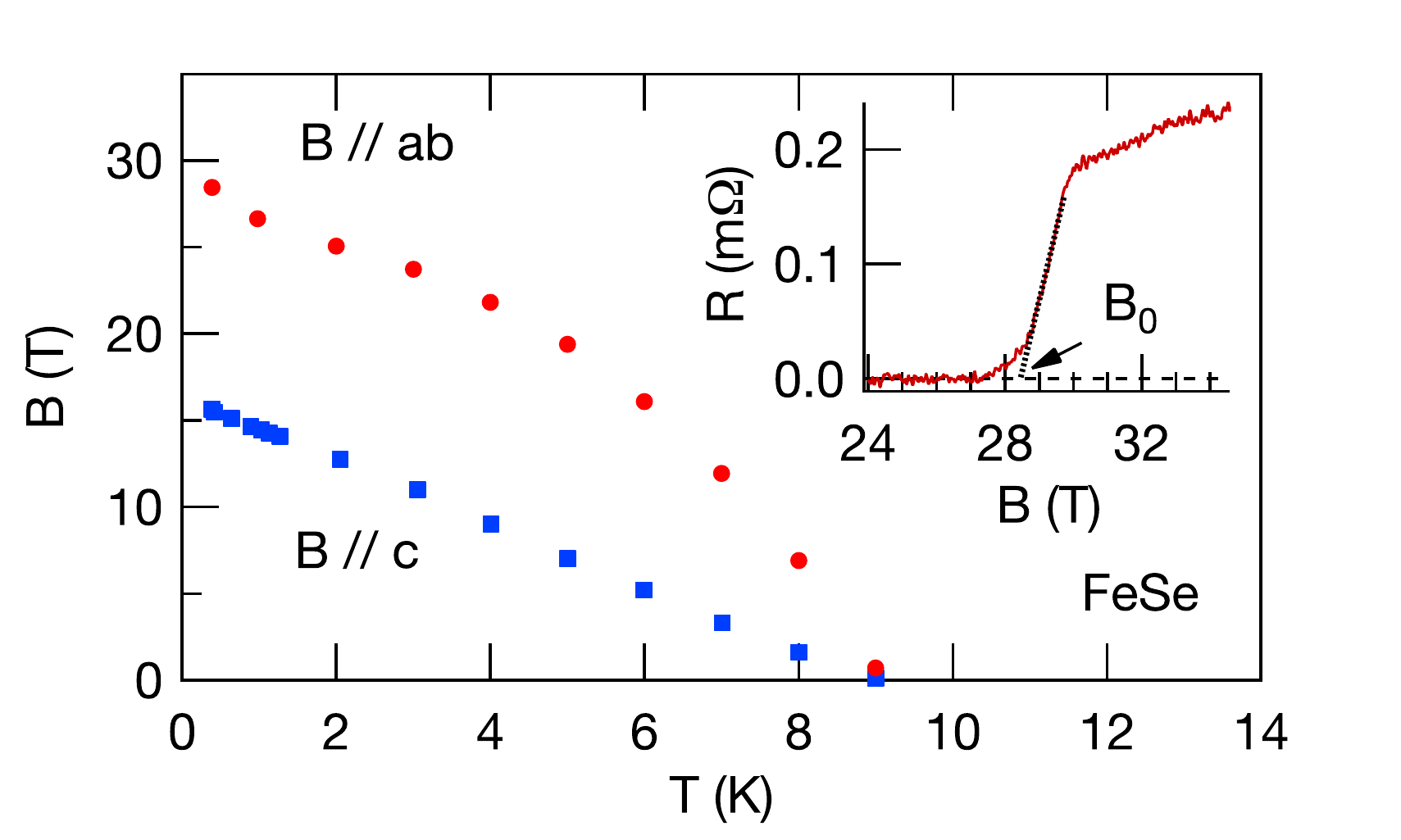}
\caption{\label{Bc2} Characteristic field $B_{0}$ in FeSe sample 2 as a function of temperature $T$. Inset: Resistance $R$ as a function of field $B$ applied parallel to the $ab$ plane at $T$ = 0.39 K.  The definition of $B_{0}$ is given.}   
\end{figure}

Figure~\ref{Bc2}  shows the temperature ($T$) dependences of the characteristic field $B_{0}$ determined from $R$ vs. $B$ curves as explained in the inset.
We assume that $B_0 \sim B_{c2}$.
We use this unconventional definition because $R(B)$ curves for $B \parallel c$ are concave in the field range just above the superconducting resistive drop [see Fig. 1(a)] and hence the usual 50 or 90\% resistive criterion for $B_{c2}$ is ambiguous.

For $B \parallel c$, $B_{0}$ increases approximately linearly with decreasing $T$. 
A similar nearly linear or even convex variation of $c$-axis $B_{c2}$ has been reported for other iron-based superconductors and has been explained by multiband effects.\cite{Hunte08Nature, Yuan09Nature, Gurevich11RPP, Terashima13PRB, Zocco13PRL}
For $B \parallel ab$, $B_{0}$ shows a tendency to saturation down to $\sim$2 K, suggesting  paramagnetic limiting, but shows an anomalous enhancement below.
A similar low-$T$ enhancement has been reported for Fe(Se, Te) \cite{Braithwaite10JPSJ, Tarantini11PRB} and (Ba,K)Fe$_2$As$_2$.\cite{Terashima13PRB}

The initial slopes are -1.6 and -6.9 T/K for $B \parallel c$ and $ab$, yielding coherence lengths $\xi$ of 1.3 and 5.7 nm for the $c$ and $ab$ directions, respectively.

The mass anisotropy $m_{\parallel c}/m_{\parallel ab}$ is estimated to be 19, which is larger than 9.5 found in LiFeAs,\cite{Kurita11JPSJ} contrary to the expectation that FeSe is more three-dimensional.

\section{ellipsoidal and hyperboloidal-surface fits}
For an ellipsoidal ($+$) or a hyperboloidal ($-$) Fermi surface $k_{ab}^2/(k_o^{ab})^2 \pm k_{c}^2/(k_o^{c})^2 = 1$, where $k_{ab}$ and $k_c$ are the $ab$ plane and $c$ axis components of the $k$ vector, respectively, the angle dependence of the frequency is given by $F(\theta) = F(0)[\cos ^2 \theta \pm (k_o^{ab}/k_o^{c})^2 \sin ^2 \theta]^{-1/2}$.
The fitting results shown in Fig. 2 are ($k_o^{ab}$ (\AA$^{-1}$), $k_o^{c}$ (\AA$^{-1}$), sign) = (0.042, 0.086, $-$), (0.078, 0.17, $-$), (0.13, 0.24, $+$), and (0.14, 0.37, $+$) for $\alpha$, $\beta$, $\gamma$, and $\delta$, respectively.

\end{document}